# Reversal of Ferroelectric Polarization by Mechanical Means

H. Lu,[1] C.-W. Bark,[2] D. Esque de los Ojos,[3] J. Alcala,[4] C.-B. Eom,[2] G. Catalan,[5,6+] and A. Gruverman[1*]

[1]*Department of Physics and Astronomy, University of Nebraska-Lincoln, NE 68588, USA*

[2]*Department of Materials Science and Engineering, University of Wisconsin-Madison, WI 53706, USA*

[3]*Department of Fluid Mechanics, GRICCA, Universitat Politecnica de Catalunya, Barcelona, Spain*

[4]*Department of Materials Science and Metallurgical Engineering, GRICCA, Universitat Politecnica de Catalunya, Barcelona, Spain*

[5]*Institut Catala de Recerca i Estudis Avançats (ICREA), Catalunya, Spain*

[6]*Centre for Investigations in Nanoscience and Nanotechnology (CIN2), CSIC-ICN, Campus de Bellaterra, Barcelona, Spain*

---

[+] Email: gustau.catalan@cin2.es

[*] Email: agruverman2@unl.edu

**Ferroelectric materials are characterized by the presence of an electric dipole that can be reversed by application of an external electric field, a feature that is exploited in ferroelectric memories. All ferroelectrics are piezoelectric, and therefore exhibit a strong intrinsic coupling between polarization and elastic deformation - a feature widely used in piezoelectric transducers and high-displacement actuators. A less explored and exploited property is flexoelectricity, i.e. the coupling between polarization and a strain gradient. Though flexoelectricity is an old concept (it was discovered in the Soviet Union almost 50 years ago [1, 2]), it is only with the advent of nanotechnology that its full potential is beginning to be realized, as gradients at the nanoscale can be much larger than at the macroscopic scale. Here, we demonstrate that the stress gradient generated by the tip of an atomic force microscope can be used to mechanically switch the polarization in the nanoscale volume of a ferroelectric film. This observation shows that pure mechanical force can be used as a dynamic tool for polarization control, enabling information processing in a new type of multiferroic high-density data storage devices where the memory bits are written mechanically and read electrically.**

One of the most distinctive features of ferroelectrics is a strong coupling between the polarization and mechanical strain that gives rise to a range of electromechanical phenomena in these materials including piezoelectric, electrostrictive and flexoelectric effects. Coupling between polarization and homogeneous in-plane strain can be exploited to control the symmetry and dielectric properties of ferroelectric thin films via strain engineering using different substrates [3, 4]. Meanwhile, the coupling between polarization and strain-gradient (flexoelectricity) offers an additional degree of freedom to control not just polarization magnitude but the polarization orientation: here, we report the unusual manifestation of the flexoelectric effect in ultrathin BaTiO$_3$ films resulting in mechanically-induced polarization reversal.

Flexoelectricity was first theoretically predicted by Kogan in 1964 [1] and experimentally demonstrated by Bursian and Zaikovskii in 1968 [2], who measured the changes in curvature of a ferroelectric film due to polarization. The same year, Scott [5] invoked a strain gradient as a parity-breaking mechanism in spectroscopic measurements. Further theoretical consideration of flexoelectricity has been provided by Tagantsev [6], and experimental measurements for perovskite ferroelectrics are due to Cross and Ma [7, 8]. Though flexoelectricity is generally weaker than piezoelectricity, gradients grow in inverse proportion to the relaxation length, so that very large flexoelectric effects can be achieved at the nanoscale [9, 10, 11, 12, 13]. Strain gradients break spatial inversion symmetry and, therefore, have polarity, so it is natural to wonder whether one can use flexoelectricity to actively switch ferroelectric polarization. If demonstrated, this would enable a new type of multiferroic memory device that is written mechanically and read electrically - the electromechanical equivalent of magnetoelectric memory devices, which use electrical writing and magnetic reading [14, 15, 16].

A large strain gradient can be realized by pressing the sharp tip of an atomic force microscope (AFM) against the sample surface, thereby causing a large stress concentration near the tip-sample contact. Previous studies of the tip-induced stress effect showed suppression of the piezoelectric response under a sufficiently high loading force [17, 18] with almost complete recovery after stress release. Polarization vector rotation due to simultaneous application of an external electrical bias and tip-induced stress has been observed in polycrystalline thin films [18, 19] - an effect that is highly unlikely in vertically polarized films since the elastic strain state is identical for domains with up and down polarizations. Abplanalp *et al* [20] have reported a higher order ferroic switching under the combined action of tip-generated compressive stress and electric field resulting in the alignment of polarization in the direction antiparallel to the applied field. In these studies, external voltage is concomitantly used to induce a change in polarization, so the role of pure mechanical stress cannot be assessed. On the other hand, recent report by Lee et al [21] showed that the flexoelectric effect in epitaxial ferroelectric films due to strain gradient could cause a strong imprint. Flexoelectricity was also invoked in the permanent reversal of polarization in polycrystalline ferroelectric films due to substrate bending [22].

To explore mechanical switching of polarization we have used epitaxial single-crystalline $BaTiO_3$ films fabricated by atomic layer controlled growth on atomically smooth (001) $SrTiO_3$ substrates with $La_{0.67}Sr_{0.33}MnO_3$ conductive buffers that serve as bottom electrodes [23, 24]. $BaTiO_3$ films with a thickness of 12 unit cells (u.c.), or approximately 4.8 nm, have been chosen so as to ensure epitaxial clamping and prevent mismatch strain relaxation [25]. Initial testing of the films by means of Piezoresponse Force Microscopy (PFM) shows that as-grown $BaTiO_3$ films are in a single-domain state with out-of-plane polarization, indicating effective screening by surface adsorbates [26].

To investigate the feasibility of polarization reversal by mechanical means, a bipolar domain pattern has been first generated using an electrically biased PFM tip. During this process, the film surface has been scanned with a tip held under +/-4 V bias that exceeds the coercive voltage for the film. The 2x2 μm² PFM images of this electrically written domain structure are shown in Figs 1a and 1b. A typical value of the contact force during conventional PFM imaging is about 30 nN. Stable and uniform

PFM amplitude signal across the domain boundary illustrates effective electric switchability of the film and strong polarization retention. This is confirmed by the local PFM spectroscopic measurements (Fig 1c), that show rather symmetric piezoelectric hysteresis loops with a coercive voltage of ~1.5 V (or coercive field of ~3 MV/cm).

The flexoelectric switching has been investigated by scanning a 1x1 μm$^2$ area of the bipolar domain pattern with the electrically grounded tip under an incrementally increasing loading force. The loading force was increased from 150 nN to 1500 nN (a corresponding change in the applied stress is estimated to be in the range from 1.5 GPa to 15 GPa). After that, a larger area of 2x2 μm$^2$ area has been imaged by conventional PFM with a low load of 30 nN (Figs 1d and 1e). It can be seen that tip-induced stress leads to a reversal of the PFM phase contrast in the left half of the image in Fig 1d, from dark to bright, indicating reversal of the polarization from up to down. Figure 1e shows a non-monotonous change in the corresponding PFM amplitude image of the flexoelectrically switched domain: initially the amplitude decreases and then, at an applied force of approximately 750 nN, starts to increase, as quantitatively illustrated in Fig 1f. This type of behavior is consistent with the polarization reversal process in PFM where the electromechanical amplitude signal passes through minimum, as in Fig 1c, due to formation of a transient polydomain structure consisting of antiparallel 180º domains with sub-resolution dimension. For reference, the PFM images of conventional electrically switched domains in the same BaTiO$_3$ film are shown in the supplementary part which exhibit exactly the same behavior under the incrementally changing electrical dc bias (see supplementary materials Fig S1). The observed mechanically induced change in the domain pattern is therefore an unambiguous demonstration of mechanically-induced polarization reversal.

There are several possible explanations for the observed mechanically-induced switching. One is piezoelectricity. When a polar ferroelectric is uniaxially compressed, the magnitude of its polarization decreases, thereby generating a surface charge density that, if sufficiently large and unscreened, could in principle switch the polarization in the opposite direction. The critical stress for this inversion is $\sigma_C = -\frac{P_S}{d_{33}}$. The saturation polarization and piezoelectric coefficients for our fully strained BaTiO$_3$ film have been calculated using the phenomenological Landau model developed by Emelyanov *et al.* [27] yielding 38 μC/cm$^2$ and 50 pm/V, respectively. This yields a critical stress of 7.6 GPa for switching, which is in the middle of the range of applied stress in our experiments. However, we can discard the piezoelectric switching mechanism for two reasons: (i) the conductive tip of our AFM is grounded during stress application, so the piezoelectric surface charge is drained away as soon as it is generated, and (ii) the piezoelectric switching mechanism should lead to the switching of domains of both polarization directions, up and down, whereas our experimental data show the switching only of the domains oriented up and not of the opposite domains.

A second hypothesis for the observed switching can infer removal of the screening adsorbates by the scanning tip: this would lead to a large depolarizing field and reduction of coercivity, at which point any asymmetry in the system, such as the difference between the bottom and top boundary conditions, could make it switch in a certain preferential direction. This hypothesis can also be ruled out based on the fact that the tip-induced mechanical switching has been observed also in BaTiO$_3$ films with top electrodes (see supplementary materials Fig S2), where adsorbates do not play any role in polarization screening.

Flexoelectricity, on the other hand, is consistent with all the experimental observations. The direction of the flexoelectric field is determined by the strain gradient, so it can switch polarization only in one direction, as observed. The flexoelectric field is tied to the strain gradient, which, being internal to the sample, is essentially insensitive to surface boundary condition. Thus, the flexoelectric mechanism is in qualitative agreement with all the experimental observations. We also can assess the feasibility of the flexoelectric switching from the quantitative point of view. The flexoelectric field is given by:

$$E_i = \frac{f_{ijkl}}{\varepsilon_i} \frac{\partial e_{kl}}{\partial x_j} \tag{1}$$

where $f_{ijkl}$ is the flexoelectric tensor, $\varepsilon_i$ the dielectric constant, $e_{jk}$ the strain and $x_l$ are the spatial coordinates of the film. The vertical component of the tip-induced flexoelectric field is therefore:

$$E_3 = \frac{f_{3311}}{\varepsilon_3}\frac{\partial e_{11}}{\partial x_3} + \frac{f_{3322}}{\varepsilon_3}\frac{\partial e_{22}}{\partial x_3} + \frac{f_{1313}}{\varepsilon_3}\frac{\partial e_{13}}{\partial x_1} + \frac{f_{2323}}{\varepsilon_3}\frac{\partial e_{23}}{\partial x_2} + \frac{f_{3333}}{\varepsilon_3}\frac{\partial e_{33}}{\partial x_3} \tag{2}$$

There are discrepancies regarding the exact value of the flexoelectric coefficients: atomistic calculations typically yield values of the order of several nC/m [28, 29], much smaller than that measured for single crystals (~10 μC/m) [30]. We use the theoretical flexoelectric tensor coefficients [28] in order to set a conservative lower bound. The dielectric constant is $\varepsilon_3 = \varepsilon_0 \varepsilon_r$, where $\varepsilon_0$ is the permittivity of vacuum and $\varepsilon_r=100$ [27].

Though the shear flexoelectric coefficients ($f_{1313}$ and $f_{2323}$) are comparable to the transverse ones ($f_{3311}$ and $f_{3322}$) [28], the actual shear strains are found to be orders of magnitude smaller than the transverse ones, so we can neglect the shear contribution. Conversely, though the longitudinal coefficient ($f_{3333}$) is an order of magnitude smaller than the others, the longitudinal strain is the biggest by an order of magnitude, so it cannot be neglected (see supplementary materials). Thus, the flexoelectric field can be approximated as:

$$E_3 = \frac{f_{3311}}{\varepsilon_3}\frac{\partial e_{11}}{\partial x_3} + \frac{f_{3322}}{\varepsilon_3}\frac{\partial e_{22}}{\partial x_3} + \frac{f_{3333}}{\varepsilon_3}\frac{\partial e_{33}}{\partial x_3} \qquad (3)$$

This is the vertical derivative of the flexoelectric potential (or flexoelectric voltage), which is given by:

$$V_3 = \frac{f_{3311}}{\varepsilon_3}e_{11} + \frac{f_{3322}}{\varepsilon_3}e_{22} + \frac{f_{3333}}{\varepsilon_3}e_{33} \qquad (4)$$

Finite element calculations were performed to model the strains $e_{11}$, $e_{22}$ and $e_{33}$ for the AFM tip radius of 10 nm and applied loading force of 1000 nN (shown in the supplementary materials, Fig S4). From these strains, a finite element map of the flexoelectric voltage has been calculated using Eq. (4) (Fig 2a). For the loading force of 1000 nN, the flexoelectric voltage generated at the surface is found to be 0.5 V. This value is comparable with the coercive voltage measured under a high loading force (see supplementary materials Fig S3) in spite of our conservative choice of flexoelectric coefficients. The electric field calculated as the derivative of the flexoelectric potential reaches values in excess of 2 MV/cm (Fig 2b) - strong enough in comparison with the typical coercive fields of ferroelectric films (several hundred kV/cm) thus validating the feasibility of the flexoelectric switching.

There are several noteworthy features of the mechanical switching: (1) it generates stable domain patterns exhibiting no relaxation within at least several hours after switching; (2) mechanically-written domain patterns are electrically erasable; (3) no damage to the sample surface due to a high loading force was observed; and (4) the mechanically-written domains are nanoscopic. These features are illustrated in Fig 3. A structure consisting of mechanically written parallel linear domains shown in Fig 3a has subsequently been transformed into a pattern in Fig 3b by electrically erasing central domain segments using a tip under a dc -3 V bias. The topographic image of the same area of the BaTiO$_3$ film (Fig 3c) acquired after this procedure does not exhibit any traces of surface deformation. Finally, Fig 3d shows an array of dot domains as small as 30 nm in size that were flexoelectrically written by abruptly changing the tip loading force between 30 nN and 1500 nN during scanning.

These results open up an alternative way to write ferroelectric memory bits using mechanical force instead of electrical bias in high-density probe-based data storage devices. Because diameter of the tip-sample contact area typically does not exceed 10 nm, the flexoelectric switching can be induced in a highly localized area allowing fabrication of a high-density domain dot pattern. Further practical advantages stem from the fact that no voltage is applied during mechanical switching, so leakage and/or dielectric breakdown are less of a problem. Once the domain pattern is written, it can be read in a non-destructive manner by a variety of local probe-based methods: by detecting the surface charge [31], by PFM imaging [32] or by measuring the electroresistive effect [33, 34]. In principle, fully mechanical write-erase operation is possible but it requires inverting the sign of the strain gradient. This can be achieved by accessing the film from the opposite size as, for example, used to be done in the double-density floppy disks of yesteryear. By converting mechanical pressure into readable information, flexoelectricity also brings up to date an even older information processing device: the AFM performing the mechanical writing operation described here has in essence become a nanoscopic typewriter.

**Acknowledgements**

G. C. and A. G. thank the Leverhulme trust for the funds that have enabled this collaboration. G. C. acknowledges financial support from grants MAT2010-10067-E and MAT2010-17771, and J. A. acknowledges support from grant MAT2011-23375 (Ministerio de Educacion, Ciencia e Innovacion). The work at University of Nebraska was supported by the Materials Research Science and Engineering Center (NSF grant DMR-0820521) and by the U.S. Department of Energy, Office of Basic Energy Sciences, Division of Materials Sciences and Engineering (DOE grant DE-SC0004876).

**Author Contributions**

A.G and G.C. conceived the idea, designed the experiment and wrote the paper. H.L. implemented experimental measurements. C.W.B. and C.B.E. fabricated the samples, E. O., J.A. and G.C. performed finite element calculations.

**Samples and Methods Summary**

Single-crystalline epitaxial BaTiO$_3$ heterostructures have been grown by pulsed laser deposition (PLD) on atomically smooth (001) SrTiO$_3$ substrates with 30-nm-thick La$_{0.67}$Sr$_{0.33}$MnO$_3$ electrode. Reflection high-energy electron diffraction (RHEED) has been used for in-situ monitoring of the layer-by-layer growth process. Before deposition, low angle miscut (<0.1°) SrTiO$_3$ substrates were etched using buffered HF acid for 90 seconds to maintain Ti-termination and then were annealed in oxygen at 1000°C for 12 hours to create atomically smooth surfaces with single-unit-cell-height steps. During deposition of all the layers substrate temperature was maintained at 680 °C with chamber oxygen pressure kept at 150 mTorr. The top SrRuO$_3$ electrode was grown at lower temperature (600 °C) to avoid intermixing. The samples were annealed at growth temperature and 1 atm oxygen pressure for 30 minutes and then cooled down to room temperature.

Polarization imaging and local switching spectroscopy has been performed using a resonant-enhanced Piezoresponse Force Microscopy (MFP-3D, Asylum Research). Conductive silicon cantilevers (DPE18/Pt, Mikromasch) have been used in this study. PFM hysteresis loops were obtained at fixed locations on the BaTiO$_3$ electrodes as a function of a dc switching pulses (250 ms) superimposed on ac modulation bias with amplitude of 0.8V$_{p-p}$ at 320 kHz. Tip contact forces have been calibrated by measuring force-distance curves.

The finite element simulations were performed for an AFM tip in contact with a BaTiO$_3$ film of 5 nm thickness attached to a semi-infinite SrTiO$_3$ substrate. The full anisotropy of the contact behavior was taken into account through the following coefficients for the stiffness matrices. For BaTiO$_3$, $C_{11}$ = 358.1 GPa, $C_{12}$ = 115.2 GPa and $C_{14}$ = 149.8 GPa [34]. For the SrTiO$_3$ substrate, $C_{11}$= 421 GPa, $C_{12}$ = 122.1 GPa, $C_{14}$ = 133.2 GPa [35]. Report on the results for a blunt punch configuration. As compared to an ideal spherical shape, a blunt punch configuration better reproduces the nanoscale surface irregularities present in a true AFM tip.

**Figure Captions**

**Figure 1.** Mechanically induced reversal of ferroelectric polarization. (a, b) PFM phase (a) and amplitude (b) images of the bidomain pattern electrically written in the BaTiO$_3$ film. (c) Single-point PFM hysteresis loops of the BaTiO$_3$ film. (d, e) PFM phase (d) and amplitude (e) images of the same area after the 1x1 μm$^2$ area in the center marked by a dashed-line frame has been scanned with the tip under an incrementally increasing loading force. The loading force was increasing in the bottom-up direction (marked by a black arrow in (e)) from 150 nN to 1500 nN. (f) PFM amplitude as a function of the loading force obtained by cross-section analysis along the white line in (e).

**Figure 2.** (a) Finite element modeling of the flexoelectric voltage generated in the film by a tip pressure of 1000 nN. (b) Variation of the flexoelectric field across the film calculated as the derivative of the flexoelectric voltage. Inset illustrates the strain gradient in experiment geometry.

**Figure 3.** Fabrication of nanoscale domain patterns by mechanical means. (a) Domain lines mechanically written in the BaTiO$_3$ film by scanning the film with a tip under a loading force of 1500 nN. (b) The same domain structure modified by electrical erasure of the mechanically written domains. Erasure has been performed by scanning the central segment with the tip under a dc -3 V bias. (c) Topographic image of the same area showing that the film surface is not affected by mechanical writing. (d) An array of flexoelectrically written dot domains illustrating a possibility of using mechanical writing for high-density data storage application.

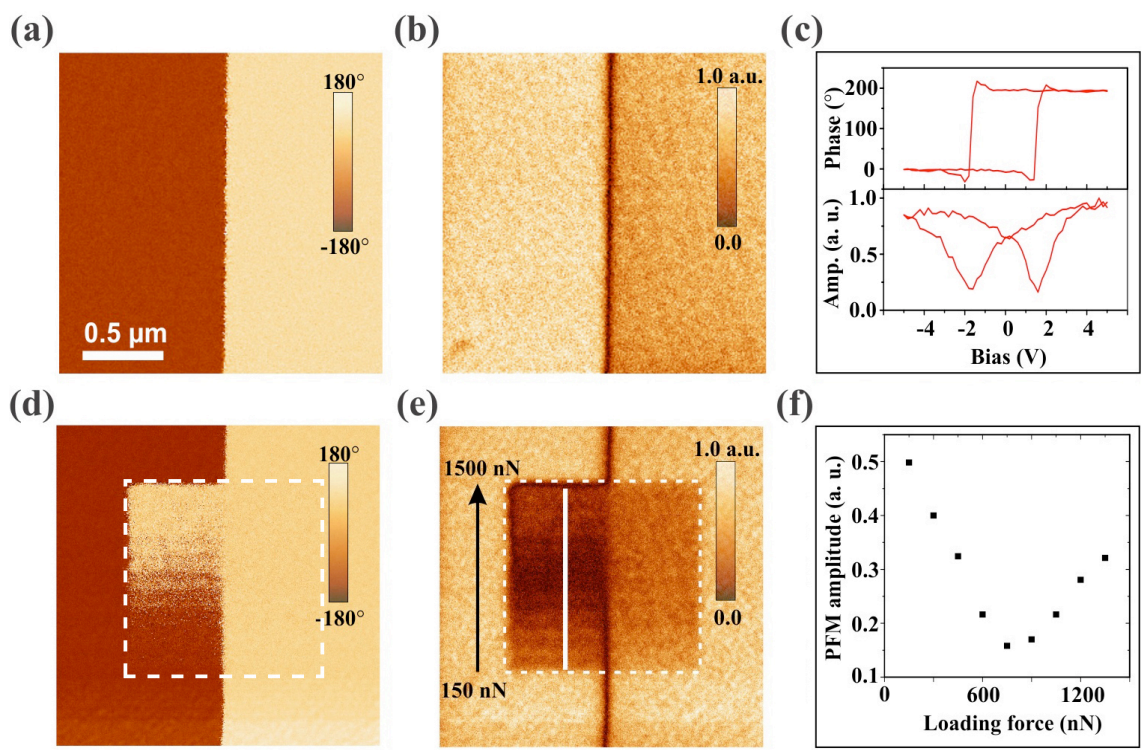

**Figure 1**

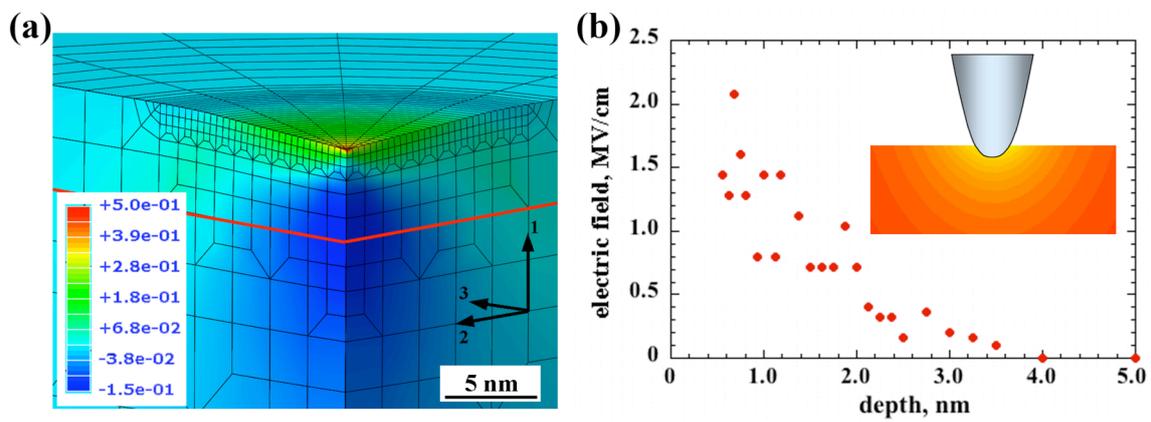

**Figure 2**

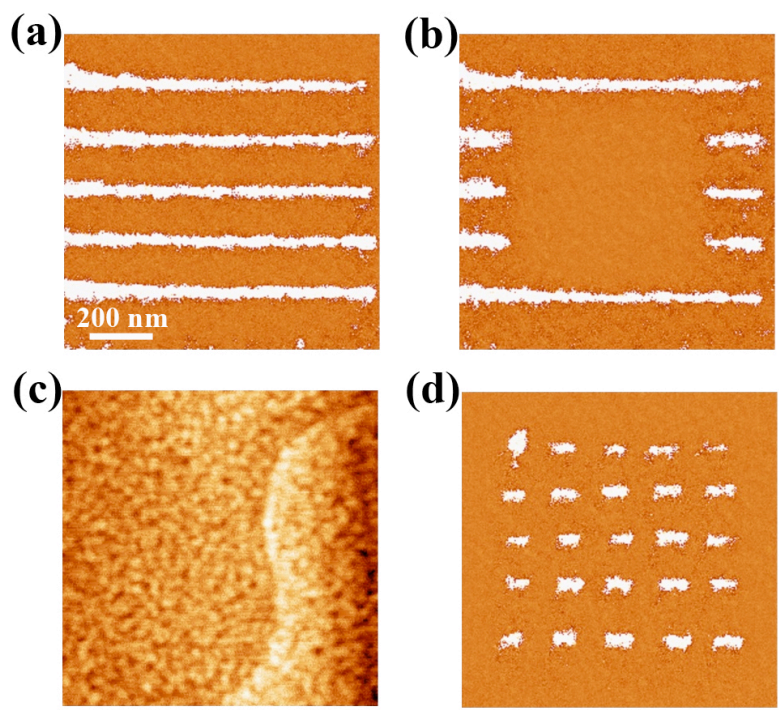

**Figure 3**

[35] Piskunov, S., Heifets, E., Eglitis, R.I., Borstel, Bulk properties and electronic structure of SrTiO3, BaTiO3, PbTiO3 perovskites: an ab inition HF/DFT study. Comp. Mater. Sci. 29, 165 (2004).

*Supplementary materials*

**Reversal of Ferroelectric Polarization by Mechanical Means**

H. Lu,[1] C.-W. Bark,[2] D. Esque de los Ojos[3], J. Alcala,[4] C.-B. Eom,[2] G. Catalan,[5,6+] and A. Gruverman[1*]

[1]*Department of Physics and Astronomy, University of Nebraska-Lincoln, NE 68588, USA*
[2]*Department of Materials Science and Engineering, University of Wisconsin-Madison, WI 53706, USA*
[3]*Department of Fluid Mechanics, GRICCA, Universitat Politecnica de Catalunya, Barcelona, Spain*
[4]*Department of Materials Science and Metallurgical Engineering, GRICCA, Universitat Politecnica de Catalunya, Barcelona, Spain*
[5]*Institut Catala de Recerca i Estudis Avançats (ICREA), Catalunya, Spain*
[6]*Centre for Investigations in Nanoscience and Nanotechnology (CIN2), CSIC-ICN, Campus de Bellaterra, Barcelona, Spain*

[+] Email: gustau.catalan@cin2.es
[*] Email: agruverman2@unl.edu

Figure S1 shows conventional PFM images of the BaTiO$_3$ film where, prior to imaging, an area in the center has been scanned with a tip under an incrementally increasing electrical dc bias. The similarity between the changes in the PFM contrast induced by conventional electric biasing (Fig S1) and those induced by mechanical loading (Fig 1(d, e) of the main text) implies the same underlying mechanism of polarization reversal in both cases.

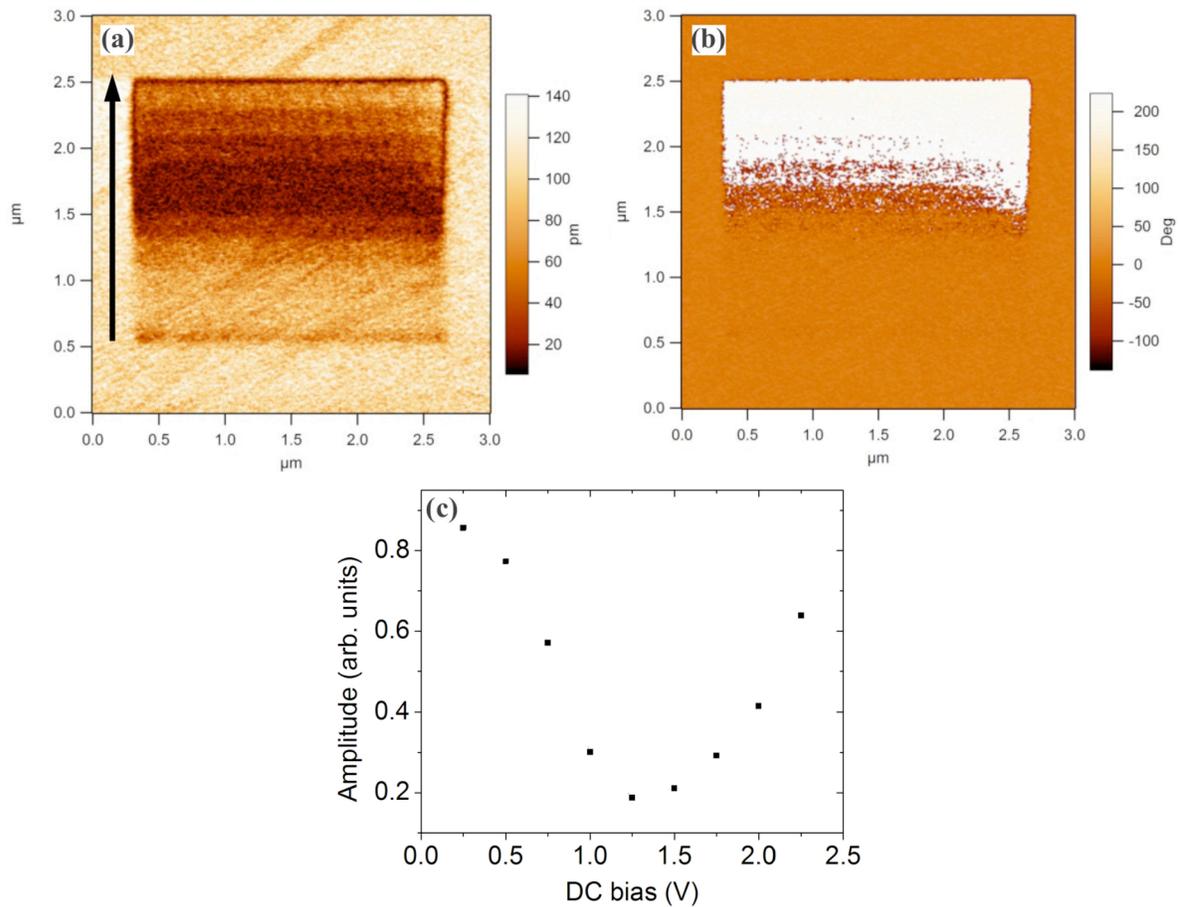

**Figure S1.** Electrically-induced reversal of ferroelectric polarization in the BaTiO$_3$ film. (a) PFM amplitude and (b) phase images of the film acquired after the central area has been scanned with the tip under an incrementally increasing dc bias. The dc bias was increasing in the bottom-up direction (marked by a black arrow in (a)). (c) PFM amplitude as a function of DC bias obtained by cross-section analysis of the poled area in (a).

The PFM amplitude contrast in Fig S1(a) gradually changes with the applied electric dc bias. Specifically, initially the PFM amplitude signal decreases and then, after reaching a minimum value at the applied bias of 1.25 V, it starts to increase (Fig S1(c)). This change in PFM amplitude is analogous to what we observed in the measurements as a function of mechanical stress (Figs 1(d) and 1(f) of the main text). A corresponding change in the PFM phase image (Fig S1(b)) shows contrast inversion at the same dc bias at which the PFM amplitude reaches its minimum, also in complete analogy with the mechanically induced phase change in Fig 1(e).

In both cases, the PFM response can be interpreted in terms of domain nucleation and formation of a polydomain structrure consisting of antiparallel 180 domains followed by emergence of a single-domain state under the increasing dc bias. Assuming that the domain size is smaller than the tip-sample contact radius, PFM will pick up a net response from a number of antiparallel 180º domains, averaging over their volume fractions. A minimum in the PFM amplitude corresponds to the situation at which there are equal fractions of opposite 180º domains.

Figure S2 illustrates the effect of the tip-induced mechanical load on the PFM signal of the 4.8-nm-thick $BaTiO_3$ film with the top $SrRuO_3$ electrode (~6.5 μm in diameter). Application of an incrementally increasing mechanical load (no dc bias is applied) leads to a stronger PFM signal as a result of mechanically-induced polarization reversal due to the flexoelectric effect as is described in the main text. Note that in the presence of the top electrodes removal of the adsorbates should not affect polarization state as the screening is mainly due to the charges in the electrodes.

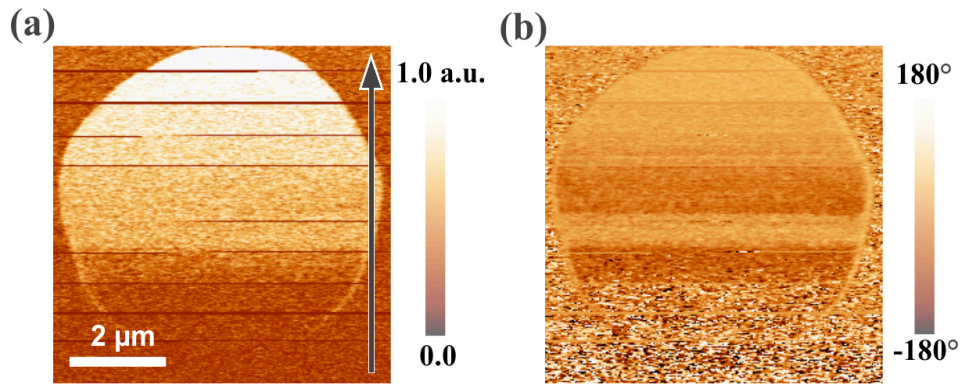

**Figure S2.** (a) PFM amplitude and (b) phase images of the 4.8-nm-thick $BaTiO_3$ film with the top $SrRuO_3$ electrode during scanning with the tip under an incrementally increasing mechanical load. The load was increasing in the bottom-up direction (marked by the arrow in (a)). Note increase in the PFM signals with the load.

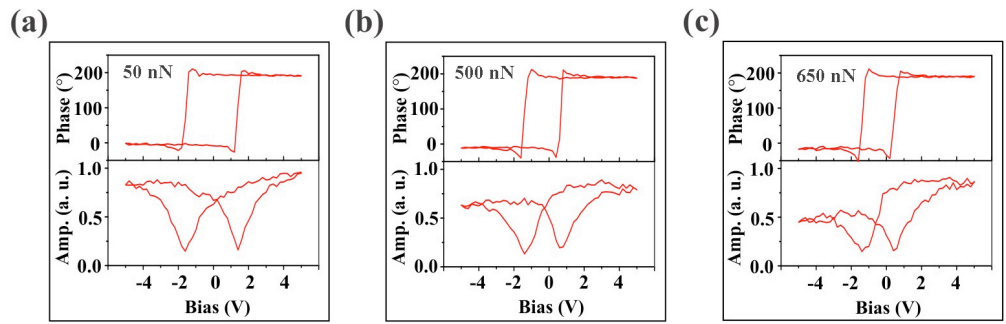

**Figure S3.** PFM hysteresis loops acquired at different loading forces. Note a shift of the loops towards the negative bias upon the load increase.

Figure S3 shows several PFM hysteresis lops acquired in the $BaTiO_3$ film at different loading forces. It can be seen that, under increasing load, the hysteresis loops become increasingly asymmetric. At sufficiently high load (<1000 nN) the film becomes unswitchable within the used voltage range, i.e. no hysteresis loop can be measured (not shown). This shows that at the high loading force the flexoelectric bias is larger than the applied voltage of the opposite polarity, supporting the flexoelectric origin of the mechanically-induced switching.

The map of the flexoelectric potential (Fig 2a of the article) was obtained by multiplying each strain field ($e_{11}$, $e_{22}$ and $e_{33}$) by their corresponding flexoelectric coefficient. The strain fields were calculated by finite element methods as explained in the experimental section, and their corresponding maps are shown in Fig S4. It can be seen that the maximum longitudinal strain $e_{33}$ (and its gradient) is bigger than the transverse ones, thus justifying its inclusion in the final calculation in spite of the smallness of the longitudinal flexoelectric coefficient.

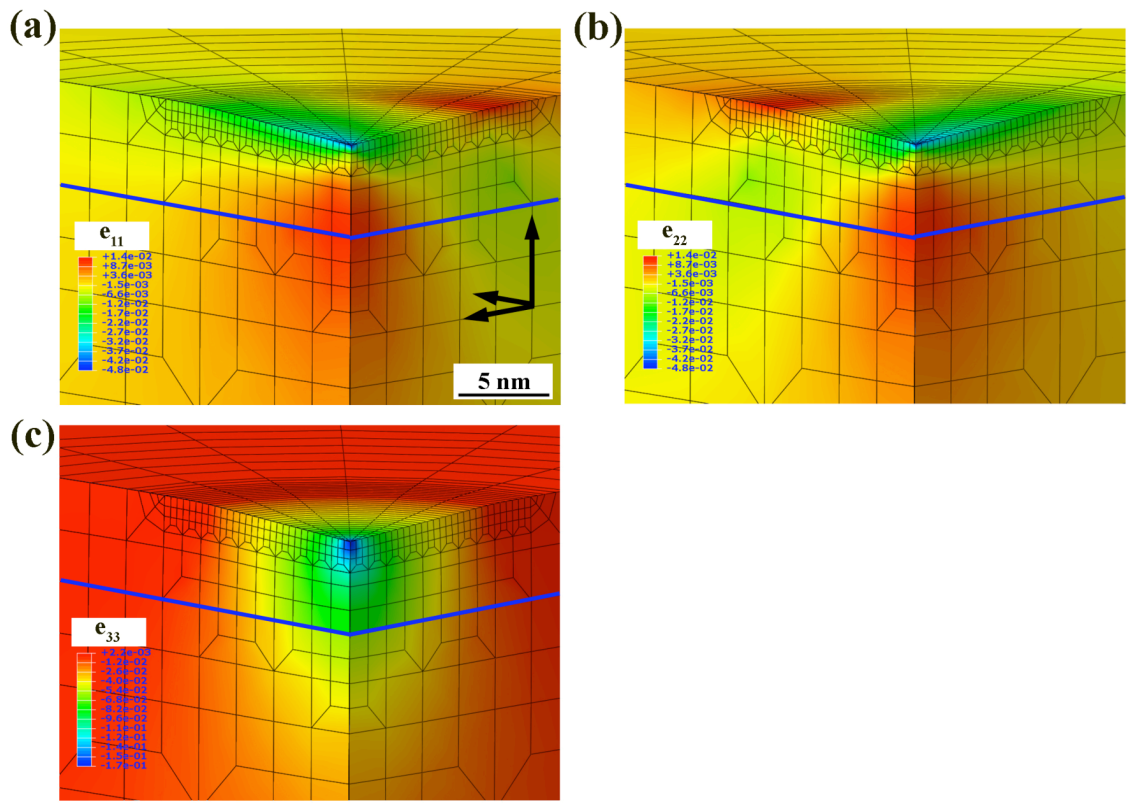

**Figure S4**. Finite element calculation of the strain fields along the x, y and z directions.